\documentclass[preprintnumbers,amsmath,amssymb]{revtex4}

 \usepackage{remreset}

\usepackage[latin1]{inputenc}

\usepackage{graphicx}
\usepackage[bf]{caption2}
\newcommand{\be}{\begin{equation}}
\newcommand{\ee}{\end{equation}}
\newcommand{\bea}{\vspace{0.25cm}\begin{eqnarray}}
\newcommand{\eea}{\end{eqnarray}}

\def\PLA{{Phys. Lett.}  A }

\def\PRL{{Phys. Rev. Lett.} }
\def\PRA{{Phys. Rev.} A }

\def\PRD{{Phys. Rev.} D }

\newlength{\defbaselineskip}
\begin{document}

\title{Interpretations of Quantum Mechanics and the measurement problem}% Force line breaks with \\

\author{Marco Genovese}
 \email{m.genovese@inrim.it}

\affiliation{%
I.N.RI.M -- Istituto Nazionale di Ricerca Metrologica\\%\textbackslash\textbackslash
Strada delle Cacce, 91 10135 Turin (Italy)%institution and/or address\\
}%
\homepage{http://www.inrim.it/~genovese/marco.html}
\date{\today}% It is always \today, today,

\begin{abstract}

We present a panoramic view on various attempts to "solve" the
problems of quantum measurement and macro-objectivation, i.e. of the
transition from a probabilistic quantum mechanic microscopic world
to a deterministic classical macroscopic world.
\end{abstract}

%%%%%%%%%%%%%%%%%%%%%%%%%%%%%%%%%%%%%%%%%%%%%%%%%%%%%
\maketitle \vskip 0.5cm Keywords: {macro-objectivation,quantum
coherence, foundations of quantum mechanics}
\section{Introduction}

Quantum Mechanics (QM) \cite{j} represents nowadays one of the
pillars of modern physics: so far a huge amount of theoretical
predictions deriving from this theory has been confirmed by very
accurate experimental data, while the theory is at the basis of a
large spectrum of researches\cite{seed} ranging from solid state
physics to cosmology, from bio-physics to particle physics.
Furthermore, in the last years the possibility of manipulating
single quantum states has fostered the development of promising
quantum technologies as quantum information (calculus,
communication, etc.), quantum metrology, quantum imaging, ...

Nevertheless, even after a pluri-decennial debate many problems
related to the foundations of this theory persist, like non-local
effects of entangled states, wave function reduction and the concept
of measurement in Quantum Mechanics, the transition from a
microscopic probabilistic world to a macroscopic deterministic world
described by classical mechanics (macro-objectivation) and so on.
Problems that, beyond their fundamental interest in basic science,
now also concern the impact of these developing technologies
\cite{QT,mat}.

It is also worth to mention that this debate expanded beyond the
physicists community, involving since its beginning philosophers
with epistemological interests, as Popper or Feyerabend, and the
development of a quantum logic \cite{QL}.

In this paper, we wish, without any pretension of being exhaustive
due to the huge published material on the subject, give a summary
introduction to the main attempts for solving the measurement
problem (and the related one of macro-objectivation) in QM. In our
purpose this should represent a summary addressed not only to
physicists, but to any scientist interested in this problematic.
Since we intend to present the main attempts to solve the problem,
we will group different similar proposals without to much attention
to nuances distinguishing among them (even if for the authors of
different proposals these "nuances" may have a large relevance).  A
large bibliography is provided that will allow the interested reader
to deepen specific interpretations or models and appreciate these
differences. Furthermore, being addressed to a general audience, we
will avoid too technical distinctions and details, preferring some
simplification for improving the readability to an absolute
precision in defining the concepts (of course when this does not
lead to errors or misunderstandings).

\section{    THE MACRO-OBJECTIVATION AND MEASUREMENT PROBLEM IN QUANTUM MECHANICS}

\subsection{ The Von Neumann chain}

One of the most characteristic properties of QM is the superposition
principle \cite{j}, i.e. the fact that a linear superposition of
states (vectors of a Hilbert space) describing the system is still a
valid state of the system. This assumption is at the basis of
interference properties and probabilistic structure of the theory
and it is a pervasive QM characteristic\cite{qs}.

When extended to many particle systems this assumption leads to
situations where the multi particle states cannot be factorized in
single particle ones, a property called entanglement. This property
leads to several counter-intuitive, paradoxical aspects of QM, as
non-locality \cite{nl}, EPR \cite{EPR} effect and so on and was
define by Schr\"odinger \cite{sch} the "the characteristic trait of
quantum mechanics".

The problem of macro-objectivation derives by the fact that
evolution equation in Standard Quantum Mechanics (SQM) \footnote{in
the following we denote by QM quantum mechanics including all its
extensions, by SQM quantum mechanics formalism without any extension
or interpretation.} is linear and thus requires that a macroscopic
system interacting with a state in a superposition becomes entangled
with it.

For example let us consider a macroscopic measurement apparatus
described by the state $| \chi _0 \rangle$ (i.e. a wave function as
complicated as necessary), which interacts with the (microscopic)
states $\vert \phi_1 \rangle $ and $ \vert \phi_2 \rangle$.

 The
interaction, representing the measurement and lasting a time
interval $\Delta t$, can be described by a linear evolution operator
$U(\Delta t)$. The results of the measurement are then

\be
\vert \chi_0 \rangle \vert \phi_1 \rangle \rightarrow
U(\Delta t) [\vert \chi_0 \rangle \vert \phi_1 \rangle ] =
\vert \chi_1 \rangle  \vert \phi_1 \rangle
\ee

and \be \vert \chi_0 \rangle \vert \phi_2 \rangle \rightarrow \vert
\chi_2 \rangle  \vert \phi_2 \rangle \ee where the states $\vert
\chi_{1(2)} \rangle$ of the measuring apparatus represent the
situations where a pointer denotes to have measured the state $\vert
\phi_{1(2)} \rangle$.

If $| \chi _0 \rangle$ interacts with the superposition state \be a
\vert \phi_1 \rangle + b \vert \phi_2 \rangle \ee because of
linearity of the evolution equation, one has \be \vert \chi_0
\rangle [a \vert \phi_1 \rangle + b \vert \phi_2 \rangle ]
\rightarrow[a \vert \chi_1 \rangle  \vert \phi_1 \rangle +b \vert
\chi_2 \rangle \vert \phi_2 \rangle ] \label{vns} \ee which is an
entangled state involving the macroscopic apparatus as well.

Of course at a macroscopic level we do not perceive anything which
can be thought as a superposition of two macroscopic situations, for
example if the measuring apparatus has a pointer that is up or down
according to if it has measured a property 1 or 2, we always observe
the pointer in one well defined position and never in a undefined
superposition of pointer up and down at the same time.

A very illuminating example of this problem was proposed by
Schr\"odinger. Let us consider a box with a cat inside. In the box
there is also a measurement apparatus that measures a property of a
quantum system, which is in a superposition state for the measured
observable. According to the value of the measurement the apparatus
open or not a poison bottle. Thus, in this case the von Neumann
chain includes the quantum system, the measuring apparatus and the
poison bottle. But at the end also the cat is involved: if the
poison has been diffused the cat dies, otherwise it survives. The
result of this analysis is therefore that we have a superposition of
cat alive and dead, which looks rather a paradoxical situation. From
this example in the literature a superposition of two macroscopic
states is usually dubbed a "Schr\"odinger cat".

Therefore,  measurements in quantum mechanics would seem to require
some process breaking the entanglement: among the possible outcomes
only one will be realized and observed in the measurement process,
i.e. a non-unitary evolution is needed. Only one state in the
superposition survives the measurement process, i.e. in the previous
example the measuring apparatus will be found or in the situation
described by $\vert \chi_1 \rangle$, with probability $|a|^2$, or in
the one described by the state $\vert \chi_2 \rangle $, with
probability $|b|^2$ and the measured state (if the measurement is
non-destructive) will be, correspondingly,  in the state $\vert
\phi_1 \rangle$ or $\vert \phi_2 \rangle$, respectively, after the
measurement process. This is called the wave function collapse.
However, this request must be justified more precisely. We have to
understand at which point of the measurement process the  collapse
occurs and how this collapse happens.

A first answer is to split the world into a macroscopic one
following classical mechanics and a microscopic one following QM, at
the moment of the measurement by a classical apparatus the wave
function collapses in one of the possible states. This is
substantially the one adopted by the Copenhagen school even if a
clear definition of what is Copenhagen interpretation is difficult
to give, since different physicists (as Bohr or Heisenberg) exposed
different points of view. For example according to Bohr classical
concepts are somehow autonomous from QM (in a way reminiscent of
Kant's transcendental arguments)\cite{bac}. However this solution,
even if perfectly useful for practical calculation of quantum
processes, is weak from a conceptual point of view since it does not
permit to identify the border between quantum and classical worlds.
How many particles should a body have for being macroscopic? What
about "macroscopic" systems as superconductors  exhibiting quantum
properties? For many reasons this answer looks to be unsatisfactory.

 Various different ideas have been considered for
 explaining/understanding decoherence at macroscopic level, without reaching for any of them
a general consensus in the physicists community. Among them (without
any purpose to be exhaustive) we can distinguish schemes where QM is
"interpreted", without modifying the formalism, and schemes where
the formalism of QM is modified or considered as incomplete. In the
firs group one can mention: the many worlds models \cite{everett},
modal interpretations \cite{modal1,modal2,modal3}, decoherence and
quantum histories schemes \cite{books,gr,old,zurek,GMH},
transactional interpretation \cite{cram}, 'informational'
interpretation \cite{zeiinf,fuchs} and many others (see for example
\cite{omn09,books4,stat,rod,par,bub,HW,lau} and Ref.s therein).

In the second group: dynamical reduction models (where a non-linear
modification of Schr\"odinger equation is introduced)
\cite{GRW,pearle,mil}, reduction by consciousness (wave function
collapse happens at observer level) \cite{wigner} or hidden variable
theories\cite{belinf,physrep}. In this last case, macrobiectivation
problem simply does not exist  since in these models the
specification of the state by using state vectors is insufficient,
there are further parameters (the hidden variables) that we ignore
for characterizing the physical situation. The physical system is
always in a well specified state (corresponding to one of the
quantum mechanical states present in the superposition) univocally
determined by the value of the hidden variables. However, it must be
noticed that for contextual theories one can attribute an objective
state only to those variables which are non-contextual.

In the following we will present some of the most interesting
attempts to solve macro-objectivation problem according to our
opinion.

\section{ Incompleteness of quantum mechanics}

A first class of schemes consider QM as incomplete: thus one has to
modify it by introducing some further element.

\subsection{Hidden variable models}

According to some authors \cite{ballen} a pure state does not
describe individual systems, but an ensemble of similarly prepared
systems: thus the formalism of QM is applicable only to groups of
similar events and not to isolated events. As, in this case, QM
predicts nothing which is relevant to a single system: the
probabilities $|a|^2$ and $|b|^2$ of finding the apparatus in the
state $\vert \chi_1 \rangle$ or in $\vert \chi_2 \rangle$ (namely
with the pointer indicating 1 or 2) merely represent the frequency
distribution of the possible measurement for an ensemble with a
given state preparation. This would strongly limit the predictivity
of QM; individual events are often met in physical investigation.

Of course, the suggestion that quantum states should refer to
ensembles of similarly prepared systems and that QM cannot describe
individual systems opens the door for hidden variables theories
\cite{belinf,physrep}. In fact, in this case is reasonable to assume
that quantum mechanics is not a complete theory, but it is  a
stochastic approximation of some deterministic theory.

This  is just the hidden variable program, mentioned in the previous
section. In this case no real entanglement exists: hidden variables
completely specify the state of the system. We do not have really
superpositions and thus the macro-objectivation problem simply does
not exist. For what concerns contextual HVT (a theory is said to be
non contextual when the value of a quantity is determined regardless
of which other quantities are simultaneously measured along with
it), like de Broglie Bohm one\cite{deb,BH}, this solution concerns
only non-contextual observables, but only non-contextual observables
can really be considered objective.

However, Bell \cite{bell} has demonstrated that a local
\footnote{i.e. a theory where the choice of measurement settings in
some place cannot affect in any way the result of a measurement
performed at a space-like (i.e. not connected with a sub-luminal
signal) distance.} HVT cannot reproduce all the results of QM and
following experiments have confirmed \cite{exp} (even if some
remaining experimental loophole \cite{santos} leaves some space for
specific models \cite{loop}) the results of QM. Nevertheless, these
tests do not concern non-local \footnote{non-locality that is anyway
such not to allow superluminal signalling.}  (as de Broglie-Bohm
\cite{BH,deb}or Nelson \cite{nlHVT} models) or Planck scale HVT
\cite{nlP,physrep} (that represent also a very interesting attempt
to reconsider the problem of quantum gravity).

Since the discussion about HVT and their tests has been the subject
of a recent review paper of the author \cite{physrep}, we address
the interested reader to this (for some more recent study on HVT see
also \cite{rHVT}). Nevertheless, it is worth noticing that these
theories, at least in principle, can be tested experimentally and
thus a future discrimination between them and SQM is envisageable
\footnote{with the possible exception of dBB and Nelson models that
are usually considered to be built to be equivalent to SQM (for some
discussion on this point see \cite{dBBdisc}).}.

\subsection{ The GRW model}

Another possible solution of the macro--objectivation problem is to
suppose a non-linear modification of evolution equation leading to
the collapse of the wave function \cite{pea,GRW,col}.

In particular we detail here a little the model that has been
proposed by Ghirardi-Rimini-Weber \cite{GRW}.

The idea consists in considering an extension of quantum mechanics
where the wave function suddenly randomly collapses according to

\be \Psi(x_1,...,x_N) j(x-x_i) / R \label{qc} \ee

where \be j(x-x_i) = A \exp [-(x-x_i)^2/ (2 a)^2] \label{j} \ee and
\be |R(x)|^2 = \int dx_1 ...  dx_N |\Psi(x_1,...,x_N) j(x-x_i)|^2
\ee is a normalisation factor.  $x_i$ is the specific coordinate of
the $i$th particle of the system (the one whose coordinate is
"fixed" by the collapse) .

The probability of the collapse is given, for each particle, by
$1/\tau$, where $\tau$ can be fixed to be $\approx 10^{15} s \approx
10^{8}$ years.

For the constant $a$, which appears in Eq. \ref{j}, GRW suggested $a
\approx 10^{-7}$ m.

Finally, the collapse centre $x$ is randomly chosen with probability
distribution $|R(x)|^2 $.

The effect of the collapse described by Eq. \ref{qc} is that one
position co-ordinate is fixed. If the state is a superposition of
different macroscopic states the system will thus collapse on one of
them (the one where particle $i$ has a position $x_i$) corresponding
to different position co-ordinate for the particles belonging to the
system, for example to different co-ordinates of the pointer of a
measuring apparatus.

The process happens with a very small probability for a single
particle (and thus does not modify QM predictions for few particles
systems), but for a macroscopic system, where the number of
particles is $N \simeq 10^{24}$, the probability  $N / \tau$ is
extremely high and the collapse is almost immediate.

\subsection{ Estensions of GRW model}

The original GRW model does not express the stochastic part of
evolution in a compact mathematical form. This has been obtained in
many other models where one introduces dynamical equations that
modify the Schr\"odinger one including a description of the wave
function collapse.

For example one can consider the dynamical equation of
ref.\cite{pearle}, which leads to the non-unitary evolution of the
state (for the sake of simplicity the unitary part is neglected):

\be \vert \Psi(t) \rangle = \exp [ - 1/ (4 \lambda t) (B(t)-2
\lambda t A)^2] \vert \Psi(0) \rangle \label{evol} \ee
$B(t)=\int
_0^t dt' w(t')$,   $w(t)$ being a white noise function and $A$ is an
opportune operator, whose eigenstates are the ones to which the
collapse occurs and $\lambda$ is a parameter ($\approx 10^{-16} $s).

Let us now suppose that the initial state $\vert \Psi(0) \rangle$
can be written as a superposition of two eigenstates $\vert a_i
\rangle$ of $A$ (corresponding to the eigenvalues $a_i$
respectively) as \be \vert \Psi(0) \rangle = a \vert a_1 \rangle +b
\vert a_2 \rangle \ee then the evolution of Eq. (\ref{evol}) would
lead to \be \vert \Psi(t) \rangle = a \exp [ - 1/ (4 \lambda t)
(B(t)-2 \lambda t a_1)^2]  \vert a_1 \rangle +b \exp [ - 1/ (4
\lambda t) (B(t)-2 \lambda t a_2)^2] \vert a_2 \rangle \ee showing
how a state evolves under a particular noise B(t). The evolution is
not unitary, so statevectors evolving under different B(t) have
different norms. States with larger norm are more likely and the
probability density for B(t) to be the actual noise is: \be P(B) =
\langle \Psi(t) \vert \Psi(t) \rangle = |a|^2 \exp [ - 1/ (2 \lambda
t) (B(t)-2 \lambda t a_1)^2]+
  |b|^2 \exp [ - 1/ (2 \lambda t) (B(t)-2 \lambda t a_2)^2]
\label{PB} \ee The most probable B(t) are thus $B(t)= 2 \lambda a_1
t$ and $B(t)= 2 \lambda a_2 t$. If the first is the actual one the
state collapses to $\vert a_1 \rangle$ with probability $|a|^2$ as
the second component is exponentially suppressed (and in practice
disappears as $t \rightarrow \infty$) as \be \vert \Psi(t) \rangle =
a   \vert a_1 \rangle +b exp [ - \lambda t( a_1- a_2)^2] \vert a_2
\rangle \ee On the other hand, if   $B(t)= 2 \lambda a_2 t$ the
collapse is to $\vert a_2 \rangle$ with probability $|b|^2$.

Thus, the collapse happens into a state or into the other according
to how the noise has fluctuated: it remains however unexplained why
the noise fluctuated that or the other way. This could be addressed
identifying the noise with some physical process (for example
gravitational fluctuations) and is demanded to a future theory.

This example represents a very simple model for a stocastic
equation. A variety of more sophisticated models have been proposed
\cite{pearle,reccol,reccol2}. Even if we will not discuss advantages
and defects of all of them, nevertheless we would like to point out
some general properties of this kind of models.

One first interesting point is that if we choose the operator $A$
(or better a class of operators generalising Eq. \ref{evol}) to be
of the form \be A(x,t) = {1 \over (\pi a^2)} \int dy N(y,t)
\exp[-(x-y)^2/(2 a^2)] \ee where x is the position and $N$ the
particles number operator, we recover a model very similar to the
original GRW.

Another interesting characteristic of all the  collapse models is
that a narrowed wave function increases its energy because of
uncertainty principle: this leads to a violation of energy
conservation. This can be checked looking to the evolution of the
average value of the Hamiltonian.

For example, for the Ref. \cite{pearle} model, this gives \be < H(t)
> = E(0) + 3/4 \lambda t n {\hbar ^2 \over 2 m a^2} \ee where
$\lambda$ and $a$ are the parameters previously defined, m is the
particle mass and n the number of particles.

For, let say, $10^{24}$ nucleons one has an energy increasing of
0.05 attoJoule/s, corresponding to a temperature increase of
$0.001^o K$ since the beginning of the universe! Even if this is a
very small amount, somehow a particle could suddenly gain a large
quantity of energy and there is some hope of detecting this
phenomenon. For example, an analysis \cite{pearle} has been done of
data collected in some experiment \cite{Ge} (originally addressed to
double $\beta$ decay search), where one has searched for X rays
which can be attributed to ionisation of one electron, which had got
energy by a collapse. The result of such an analysis is a limit on
$\tau$, which indicates that electrons must collapse much less
rapidly than protons (whose collapse value is fixed in the model by
the request of having a correct wave function reduction for
macroscopic bodies). This result has been related to the fact that
the electron mass is a factor 1836 smaller than the proton one and
has led to speculations about the fact that collapse could be
related to quantum gravity \cite{pearle}, as a mass dependent
coupling would seem to point out (for possible relations between
gravity and wave function collapse, see also \cite{grav}).

On the other hand, experimental tests of these models based on
interference concerning mesoscopic objects are much more difficult
to be realised.

Finally, one can also notice that the collapse is a non-local
process: an entangled state collapse could also instantly involve
two very far components of the system, however no faster-than-light
information can be transmitted using these non-local effects.

Nevertheless, the definition of "instantly" would require a
preferred frame where the collapse happens. It has been demonstrated
that this would not lead to any experimental result in disagreement
with special relativity predictions \cite{AA}. However, it looks
rather peculiar that the collapse of entangled states is such not to
permit any faster than light transmission and at the same time it
carries along non relativistically invariant phenomena. A
relativistic collapse model would be rather desirable.

Some attempts in this sense have been done \cite{pearle,reccol2}.
For example, in Ref. \cite{rel} the operator $A$ in Eq. \ref{evol}
is substituted by a scalar field. The collapse then works as
follows: a fermion in a certain superposition is entangled with the
scalar field too. The "noise" causes the collapse of the scalar
field and this involves the fermion. Incidentally, one can notice
that in this case which detection triggers the collapse and where
and when the collapse takes place become frame-dependent, but of
course these are not measurable properties. The real problem of the
model is that the collapse originates an infinite increasing of
energy (due to creation of scalar particles), which cannot be
eliminated. Some progress has been obtained
\cite{pearle,reccol2}(e.g. by introducing tachyonic fields), but a
satisfactory solution is still missing.

In conclusion, it is worth to emphasize that, as for HVT theories,
also in this case a final answer on these models will come from
experimental tests of them.

\subsection{ Reduction by consciousness}

Somehow also the interpretation of a reduction by consciousness of
Wigner\cite{wigner} can be considered as a scheme where QM is not
complete, indeed  consciousness  acquires an extra-physical role and
cannot be described by the theory.

In little more detail the Wigner's argument:
 He analysed the von Neumann sequence of
Eq. \ref{vns}, supposing that at the end a friend of his looks at
the experimental apparatus. In principle also Wigner's friend is
described by a very complex wave function, initially $\vert F_0
\rangle$. After observation one should have the entangled state \be
  \vert \chi_1 \rangle  \vert \phi_1 \rangle \vert F_1 \rangle
+ \vert \chi_2 \rangle  \vert \phi_2 \rangle \vert F_2 \rangle
\ee
where now even the Wigner's friend is in a superposition.
Then Wigner asks his friend about the result: he is sure of obtaining a well precise answer. What does it means this? Wigner cannot assume (if he does not want to assume an extreme solipsistic attitude) that  his question  causes the collapse of his friend wave function, then the collapse must be happened at some point of the von Neumann sequence. But which is the more distinctive point of the sequence? According to Wigner's answer  this point is when a perception happens. Then the sequence is interrupted when his friend observes the apparatus: it is perception to cause the collapse.

Of course, such a point of view considers the mind out of physical
world, than can be considered a weak point. Furthermore, what about
the universe evolution? The universe is remained in an extremely
complex entangled state up to when the first jelly-fish had a first
foggy perception of it? Or it has been bound to wait for the
transition between homo erectus and homo sapiens?

\section{Collapse-free approaches}

In these approaches the main idea is that the mathematical formalism
of QM is sufficient as it stands, no  changes have to be added to
it.

\subsection{The many worlds models}

According to this interpretation, due to Everett \cite{everett}, one
supposes that every quantum possibility realises even at a
macroscopic level, but in different non-comunicating universes: thus
no interference can be observed for macroscopic bodies.

Another similar hypothesis suggests that the splitting happens at
the level of mind \cite{mm}. One has different minds with different
perceptions corresponding to different component of the state. Of
course the minds of different observers must be correlated in order
to observe the same result.

Deutsch \cite{d1} has supported many world interpretation on the
basis that it gives an explanation of advantages of quantum
computation as a "parallel" calculation in different worlds;
furthermore he claims, against current opinion, that Everett scheme
has empirical differences with Standard QM \cite{d2} (as
"superposition of distinct states of consciousness"), a position
that received various critics \cite{j}.Indeed, for of the other
authors testing such a hypothesis is impossible.

Although it is surely charming and solves someway the
macro-objectivation problem, however Occam's razor, asserting that
we have to refuse a theory that introduced unnecessary elements,
could apply to it: indeed, in many worlds interpretation one
introduces the hypothesis of a continuous generation of infinite
splitting worlds without  a real necessity of doing it (other
possible explanations are available). Furthermore, the splitting
should concern every measurement process only, but not other
processes where interference appears. However, the distinction
between the two processes is not always evident.

Also, the problem of how the basis problem (i.e. how to choose in
which basis the splitting of worlds happens) is still under
discussion \cite{stapp}.

For some recent works concerning many worlds interpretation see
Ref.\cite{mwn}.

\subsection{ Decoherence}

A different, albeit connected, point of view is the one known as
quantum decoherence.

The starting point of this interpretation is considering how one can perform  a
measurement showing a macroscopic superposition.

For example, let us consider a system composed of two subsystems
dubbed $A$ and $S$.

We suppose that in $A$ one can perform measurements only on
compatible variables corresponding to different eigenspaces $\it
A_k$. Then it exists, as one could show, an operator $T$ that has
different eigenvalues $t_k$ for each eigenspace  $\it A_k$ (where $
P_k^A$ will denote the projector on this eigenspace). Every function
on $A$ can be written as a function $f(T)$ of $T$.

Let us now measure the mean value of the operator $O^S f(T)$, where $O^S$ acts on the subspace $S$ and $ f(T) $ on A. In the following $I^S$,$I^A$ are the identity operators on the subspaces $S,A$.

Then \be \langle \Psi \vert O^S f(T) \vert \Psi \rangle = \langle
\Psi \vert O^S f(T) I^S \otimes I^A \vert \Psi \rangle = \langle
\Psi \vert O^S f(T) \sum_{i,k} P_i^S P_k^A \vert \Psi \rangle =
\sum_k f(t_k) \langle \Psi \vert O^S P_k^A \vert \Psi \rangle
\label{macropuro} \ee

Let us now consider a statistical mixture of the normalised states
$ {P_k^A \vert \Psi \rangle \over \vert \vert P_k^A \vert \Psi \rangle \vert \vert }$ with weights  $p_k=\vert \vert P_k^A \vert \Psi \rangle \vert \vert ^2$,
then
\be
\langle O^S f(T) \rangle = \sum_k p_k \langle \Psi \vert P_k^A O^S f(T)  P_k^A \vert \Psi \rangle = \sum_k f(t_k) \langle \Psi \vert O^S (P_k^A)^2 \vert \Psi
= \sum_k f(t_k) \langle \Psi \vert O^S P_k^A \vert \Psi \rangle
\ee
which cannot be distinguished by the pure state result of Eq. \ref{macropuro}.

The result of this analysis is that if we are bound to measure only compatible observables for a subsystem, then the pure state $\vert \Psi \rangle$
cannot be distinguished by a statistical mixture.

Furthermore, this implies that we cannot neglect (or limit our measurements to compatible variables) any of the constituents of the von Neumann sequence, if we
want to distinguish a pure state from a statistical mixture.

The idea of quantum decoherence \cite{books,old,gr,zurek,GMH,de}  is
that the interaction with environment makes practically impossible
to identify interference for macroscopic systems  as a huge amount
of subsystems are rapidly involved and therefore considering all the
constituents of the von Neumann sequence becomes practically
impossible (somehow related to this scheme are the models based on
master equations \cite{me}).
 Performing correlation measurements on  a macroscopic system is "de facto" impossible and thus one cannot show "de facto" a
  macroscopic superposition. This can be restated by asserting that
  after interaction with environment a pure state is transformed in
 statistical mixture when environment degrees of freedom are traced
 out. e.g. the state of Eq.4 (where now the states $|\chi \rangle$ represent the environment), when considering orthogonality of
 environment states, will be described by:
 \be
 \rho^{red}=Tr^{Env} \rho^{system+Env}= |a|^2\vert \phi_1 \rangle \langle \phi_1 \vert+ |b|^2 \vert \phi_2
 \rangle \langle \phi_2
 \vert
\label{dec} \ee where \be \rho^{system+Env} = |a|^2\vert \phi_1
\rangle \langle \phi_1  \vert \bigotimes \vert \chi_1 \rangle
\langle \chi_1 \vert+ |b|^2 \vert \phi_2
 \rangle  \langle \phi_2 \vert \bigotimes \vert \chi_2
 \rangle \langle \chi_2 \vert
\label{decop}
 \ee
 However, Bell objected that in any case this leads only to a
{\it valid for all practical purposes theory}, which however does
not solve quantum measurement problem definitively. One could always
suppose to be able to prepare a very smart experiment which would
permit to show macroscopic superpositions.

The answer of decoherence scheme supporters is the attempt of
showing that such an experiment cannot be even envisaged, for it
would require either an infinite components apparatus or an infinite
measurement time. Many models have been studied for supporting this
statement, but no general prove of it has yet been found.

Another objection \cite{ghiob}  is that within QM the correspondence
between statistical ensembles and statistical operators is
infinitely many to one. Thus, even when accepting that the
statistical operator to be used is the one of Eq.\ref{decop}, there
is no reason to interpret it as describing the statistical ensemble
Eq.\ref{dec}.

For some recent development of decoherence schemes, as Quantum
Darwinism (i.e. the redundant recording of information about the
preferred states of a decohering system by its environment), see
\cite{decrec}.

\subsection{ Quantum Histories}

Decoherence models are  related to the quantum histories formulation
of Quantum Mechanics \cite{gr,books}, which somehow try to give a
more precise description of  the process of measurement in this
framework. The same objections, just quoted at the end of previous
subsection, also pertain this approach.

This formulation starts by the hypothesis that measurements have to be treated as every other interaction.

The various properties will be specified by a projector operator $P_k$, e.g. if the particle has spin in the direction z, this property is specified by the projector on the eigenfunction corresponding to the spin in the direction z (for the sake of generality, in the following $P_k$ is going to be considered as a generalised projector, projecting in a certain set of  eigenstates corresponding to same interval of the eigenvalues).

Considering a sequence of temporal instant $t_1,t_2, ...t_n$ and
making some precise assertions about the properties of the physical
system (which must be isolated) at the various instants (for example
the particle has spin in the direction z etc.), a quantum history is
then a sequence of statements like: \bea P_{k_n} exp[-i/\hbar H
(t_n-t_{n-1})] P_{k_{n-1}}  \cdot exp[-i/\hbar H (t_{n-1}-t_{n-2})]
... P_{k_1} exp[-i/\hbar H (t_1-t_{0})] \vert \Psi(t_0) \rangle \eea

where  $ \vert \Psi(t_0) \rangle  $ specifies the initial state.

Projectors $P_{k_i}$ corresponding to the same set, denoted by the
suffix $i$, are alternative and exhaustive, $P_{k_i} P_{k'_i} =
\delta_{k_i,k'_i} P_{k_i}$ and $ \sum_{k_i} P_{k_i} = 1 $. The
probability associated at each history is then

\bea P[t_n,k_n; ... t_1,k_1]= \vert \vert P_{k_n} exp[{-i \over
\hbar}H (t_n-t_{n-1})] P_{k_{n-1}}  exp[{-i \over \hbar}
H(t_{n-1}-t_{n-2})] ... P_{k_1} exp[{-i \over \hbar} (t_1-t_{0})]
\vert \Psi(t_0) \rangle \vert \vert ^2 \label{pp} \eea

Let us emphasise again that  making an assertion about the system at
a certain time $t_n$ does not require, in this formulation, having
performed a measurement. The probabilities Eq. \ref{pp} refer to the
objective fact that the system has the indicated properties ($k_1,
... ,k_n$) at the different instants $t_1, ... ,t_n$.

However, it is evident that we cannot include all the possible
histories together, otherwise the laws of probability  would not be
respected. As a simple example let us consider the very simple
history: \be P_{k_1} exp[-i/\hbar H (t_1-t_{0})] \vert \Psi(t_0)
\rangle \ee

If we sum on all the possible values of the index $k_1$ then $
\sum_{k_1} P_{k_1} =1 $ and thus the sum over the probabilities
concerning the different histories where the index $k_1$ is varied
is 1: \bea \sum_{k_1} P[t_1,k_1]= \sum_{k_1} \vert \vert P_{k_1}
exp[-i/\hbar H (t_1-t_{0})]\vert \Psi(t_0) \rangle \vert \vert ^2 =1
\eea This simply means that among all the various possible
alternatives, only one is realised.

But we could have chosen instead of the property $k_1$ another non compatible property. For example if $k_1$ is the z spin component we could  have chosen the x spin component. In this case the sum of the two families of histories would have probability 2! It is evident that we cannot consider all the histories together, but we have to select a subsample of "compatible" histories.

One can show that the necessary condition for two histories for
being compatible is that the so-called decoherence functional \bea
\langle \Psi(t_0) \vert exp[i/\hbar H (t_1-t_{0})] P_{j_1}
exp[i/\hbar H (t_2-t_{1})] P_{j_{2}} exp[i/\hbar H (t_{3}-t_{2})]
... exp[i/\hbar H (t_n-t_{n-1})]  P_{j_n} \cr P_{k_n} exp[-i/\hbar H
(t_n-t_{n-1})] P_{k_{n-1}} exp[-i/\hbar H (t_{n-1}-t_{n-2})] ...
P_{k_1} exp[-i/\hbar H (t_1-t_{0})] \vert \Psi(t_0) \rangle \eea is
zero as soon as at least one of the indexes $j_i$ differs from the
corresponding index $k_i$. This is substantially the complementarity
principle: in quantum mechanics we cannot make statements on the
value of complementary observables (as position and momentum) at the
same time.

If we consider a measurement apparatus, $A$, then the history
concerns properties of both the quantum system, $S$, and the
apparatus itself. If we are interested into the properties of the
subsystem $S$ only, the projectors appearing in the history concern
the  Hilbert space pertaining $S$ only.

A set of histories $P_{k_n} ... P_{k_1}$ determines an evolution of the initial state $\vert \Psi(t_0) \rangle$
of the form (where we sum over a subset of the possible values of $k_1$, corresponding to the ones included in the
specified history, which now we assume, more generally,  to be specified by a set of possible values of the observable):
\be
 \vert \Psi(t_1) \rangle = \sum_{k_1}   P_{k_1} exp[-i/\hbar H (t_1-t_{0}) \vert \Psi(t_0) \rangle =  \sum_{k_1} \sum_{a \varepsilon k_1} \vert \phi_a \rangle \vert \chi_a \rangle
\ee
where we have written $P_{k_1} = \sum_a \vert \phi_a \rangle \langle \phi_a \vert $ (we consider $P_{k_1} $ as
 projecting in a certain set  denoted by the index $a$ and $\vert \phi_a \rangle$ denotes the eigenfunctions basis of an observable in $S$).
  $\vert \chi_a \rangle =\langle \phi_a \vert  exp[-i/\hbar H (t_1-t_{0}) \vert \Psi(t_0) \rangle $ is a state of the Hilbert space $A$ and is
  dubbed the $A$ state relative to the $S$ state $\vert \phi_a \rangle$.
Proceeding this way, at $t=t_n$ we have:
\be
 \vert \Psi(t_n) \rangle =\sum_{k_1, ..., k_n} [\sum_{a_i \varepsilon k_i} \vert \phi_{a_i} \rangle \vert \chi_{a_i}^{k_1,...k_n} \rangle ]
\ee where inside brackets is the branch specified by indexes $ k_1,
..., k_n$ (corresponding to the observable considered at times $
t_1, ..., t_n$).

Thus, if $\vert \chi_{a_i}^{k_1,...k_n} \rangle$ with different apices are orthogonal, we have decoherent histories, namely compatible each other.
If the system is such that the $S$ states
$\vert \phi_a \rangle$ are correlated to orthogonal states   $\vert \chi_{a_i}^{k_1,...k_n} \rangle$, we have decoherence.

Up to here the histories formalism is just another way of
formulating QM, without any solution of the macro-objectivation
problem. However, the last point gives some clues how to proceed in
order to give a solution of this problem: if the system becomes
entangled with orthogonal states, decoherence appears. Nevertheless,
the effective decoherence phenomenon  is discussed in different ways
by different authors.

Omnes and others use the histories together with the idea of
environment decoherence \cite{books}, described in the previous
paragraph, and relate the macro-objectivation to the practical
orthogonality of macroscopic states $\vert \chi_{a_i}^{k_1,...k_n}
\rangle$. More in detail they argue that the evolution of the
composed system $S$, quantum system, plus $A$, apparatus, is such
that starting a certain point it exists an observable such that its
specifications establish a set of decoherent histories. The
deterministic behaviour of the "classical" state $A$ is due to the
fact that among these histories, one has almost unit probability.

For Ref.\cite{zurek} quantum histories must be coupled to the many
worlds interpretation: different incompatible histories realise in
different worlds.

Finally, for Gell-Mann and Hartle \cite{GMH} one must limit
histories to the whole universe: simplifying a bit, decoherence is
due to the fact that it is clearly impossible to have evidences of
superpositions of different states of the universe.

For some recent study in this framework see \cite{cra}.

\subsection{Informational interpretation}

In the recent years emerged, largely motivated by quantum
information studies \cite{QT}, a new approach based on the idea that
quantum mechanics must be considered as a theory about
"information", which is more fundamental that the concept of
"substance".

As the former interpretations, but even more, also in this case
there is not a univocal interpretation, but different authors inside
this framework emphasize more or less some specific point.

For example in "bayesian interpretation" is suggested that "a
quantum state is specifically and only a mathematical symbol for
capturing a set of beliefs or gambling commitments " \cite{fuchs},
i.e. following bayesian subjectivist interpretation of (quantum)
probabilities, one reaches a subjective interpretation of the
quantum state.

For Ref. \cite{svozil} information must be considered as the
fundamental basic entity, "the concept of a many-to-one state
reduction is not a fundamental one but results from the practical
impossibility to reconstruct the original state after the
measurement". Thus, somehow this approach stops to attribute an
ontology to the theory, it states that that QM gives rules about the
information one can have, not on the objects themselves.

Similar ideas have been expressed in Ref.s \cite{info} as well.

Anyway, this approach relating to the idea that physics can be
reduced to information contains several drawbacks and has been
criticized by many authors \cite{hag,gg,j}.

\subsection{The modal interpretations}

According to their proposers \cite{modal,d2007} modal
interpretations have the ambition to construe quantum mechanics as
an objective, man-independent description of physical reality. This
is achieved by stating that the relation between the formalism of
quantum theory and physical reality is to be taken as probabilistic,
i.e. the quantum formalism does not describe what actually is the
case in the physical world, but rather provides us with a list of
possibilities and their probabilities. As stated by Dieks
\cite{d2007} "the state in Hilbert space is about possibilities,
about what may be the case, about modalities".

Let us consider the bi-ortho-normal decomposition (whose existence
is guaranteed by a theorem of Von Neumann) for a composed system $A
- S$ \be \vert \Psi \rangle = \sum_{i,a} \sqrt {(p_i)} \vert
\phi_i^S \rangle \vert \chi_i^A \rangle \label{bio} \ee where the
chosen states are such that \be \langle \phi_i^S \vert \phi_j^S
\rangle = \delta_{i,j} \ee and \be \langle \chi_a^A \vert \chi_b^A
\rangle = \delta_{a,b} \ee

In modal interpretation the mathematical state represents situations
with definite physical properties even if this state is a
superposition of eigenstates of the corresponding observables,
definite values are assigned to the observables associated to the
projectors $\vert \phi_i^S \rangle \langle \phi_i^S \vert$. The
physical situation corresponding to the mathematical description of
Eq. \ref{bio} is that the partial system associated with Hilbert
space $S$, taken by itself, possesses exactly one of the properties
associated with the states $\vert \phi_i^S \rangle$. The distinct
terms in Eq. \ref{bio} are treated as distinct branches, but one is
always speaking of only one physical system at variance with
many-world interpretation, where one has proliferation of systems.
It is a fundamental property of this interpretation that a physical
system always possesses exactly one of the possible values singled
out by the form (\ref{bio}), even if they are not necessarily
characterised by classical properties as position or momentum. For
example for an isolated single particle system in the state $\vert
\phi \rangle$, the applicable property would correspond, with
probability 1, to the projector $\vert \phi \rangle \langle \phi
\vert $.

Therefore, the weight $p_i$ assumes the meaning of the fraction of
the members of the set, which are in the factorised state $\vert
\phi_i^S \rangle \vert \chi_i^A \rangle$. Wether we consider a
quantum system $S$ and a measuring apparatus  $A$, we have  that
modal interpretation states that a fraction $p_i$ of apparatuses
will be found in the well defined situation described by the state
$\vert \chi_i^A \rangle$.

However, it must be noticed that in general the choice of the
subsystems  $A$ and $S$ is not univocal. If we have a quantum system
in interaction with a measurement apparatus, it is rather obvious to
consider the quantum system as $S$ and the apparatus as $A$. But if
we consider, for example, a system of three distinguishable spin
$1/2$ particles (for example three quarks of different flavour), we
could consider particle 1 and 2 to form the system $A$ (for example
a singlet) and  particle 3 to form the system $S$. In this case we
would have definite properties for the system particle 1+2 and for
particle 3. On the other hand, we could have considered particle 1
and 3 to form the system $A$ (for example a singlet) and  particle 2
to form the system $S$. In this case we would have definite
properties for the system particle 1+3 and for particle 2.
Therefore, we must be cautious in making statements about the
objective properties belonging to a system, one cannot use
traditional logic for making assumptions about properties of the
subsystems: a new logic must be introduced (the so-called modal
logic).

From the fact that the assignment of properties depends on the
choice of $A$ and derives that the assignment of property is context
dependent: the partial system associated with $S$ is assigned of a
specific property {\it in the context of its interaction with the
environment represented by $A$}.

Furthermore, a pure state is supposed to evolve according
Schr\"odinger equation as a pure state, independently of the
statements done about its properties at a certain time. But it
remains unclear why it does not behave as a statistical mixture, if
we reinterpret the bi-orthogonal decomposition, in the modal
interpretation, as an inhomogeneous set. The modal interpretation,
somehow, does not seem to offer a real objective existence to the
properties of the system: the hypothetical subsystems seem to have
only a conceptual and not a real status. Indeed, to take properties
as existing in actuality would lead to some sort of hidden variable
theory \cite{b}.

\subsection{Relational Quantum Mechanics}

In Relational Quantum Mechanics \cite{rov,vFr} there are no observer
independent states, nor observer independent values of physical
quantities. The states are relative to the observer, which has not
in principle the connotation of consciousness since any system can
provide a frame of reference relative to which states and values are
assigned. All observer are assumed to be equivalent. Similar to the
recent developments described in subsection IV-D, Relational Quantum
Mechanics describes information that one system can have about
another.

As an example, let us imagine to have two observers A and B. A
measures a dichotomic observable O on a system S, registering 1 and,
thus, assigning the state $| O,1 \rangle $.

B has the information that the measurement is taking place, and
describes the state of A + S as $\alpha | O,1 \rangle \bigotimes |
A,1 \rangle + \beta | O,0 \rangle \bigotimes | A,0 \rangle  $ where
$| A,0 \rangle$, $| A,a \rangle$ are the pointer reading states. If
B does not make a measurement it has to assign to the observable O
of S  $0$ with probability $|\beta|^2$ and  $1$ with probability
$|\alpha|^2$. Therefore, A and B assign different states to S.

A tentative answer to many questions that one can poses about this
interpretation may be found in \cite{vFr}.

\subsection{Further interpretations}

Beyond the most discussed models and interpretations presented in
the previous sections, several others have appeared with a more
limited success.

Among them one can mention:

- \textbf{Transational Interpretation }\cite{cram}: here a retarded
offer wave (OW) $\Psi$, formally corresponding to the usual quantum
state vector, is emitted by a source. Then, depending on the
experimental arrangement, components of the OW may be absorbed by
one or more absorbers, each of them responding by sending an
advanced (time-reversed) confirmation wave (CW) $\Psi^*$, which
travels back to the emitter. When there are N such CW responses,
there are N incipient transactions in the form of OW/CW
superpositions. The Echo amplitude $\Psi \Psi^*$ at the locus of the
emitter equals the Born probability. The exchange repeats until the
exchange quantities (as energy) satisfy the quantum boundary
conditions of the system.

- \textbf{Emergence of classical world from quantum one}: Laughlin
in Ref.\cite{lau} attributes the emergence of classical from the
quantum world as deriving from the complexity (intended as a term
denoting systems whose properties cannot be derived by the
properties of their constituent) of measurement apparatuses.
Nevertheless this "answer" may look somehow a tautology, answering
to the question about why a classical apparatus does not show a
quantum behaviour with "it is too complex for showing it". However,
it is worth to mention here that recently Omnes proposed a similar
idea, i.e. that a "non-unitary" evolution can be introduced when the
"organization" of the macroscopic measuring apparatus is considered,
even if a wave function collapse model was not yet proposed
\cite{omn09}, but demanded to future developments.

- \textbf{Belavkin's scheme}: here the collapse of the wave function
is the stochastic result of a deterministic  unitary evolution when
an interaction between a  quantum and a classical system happens
\cite{bela}, classicality being related to superselection rules
stating that not all the observables are actually measurable
(quantum superposition of certain states cannot be detected).

\section{Conclusions}

In this paper we have tried to present a panoramic view on various
attempts to "solve" the problems  of quantum measurement and
macro-objectivation, i.e. of the transition from a probabilistic
quantum mechanic microscopic world to a deterministic classical
macroscopic world.

We have tried to be as unbiased as possible, presenting advantages
and disadvantages of the various proposals and leaving to the reader
to form (eventually using the large bibliography) an opinion on
them. Of course a complete objectivity and, even more, exhaustivity
in treating such arguments is impossible: we do apologize with
authors that  either are not satisfied of the description of their
schemes or have not been quoted (properly).

Nevertheless, we hope this paper will represent a useful tool for
who wants to approach this kind of studies.

%%%%%%%%%%%%%%%%%%%%%%%%%%%%%%%%%%%%%%%%%%%%%%%

\end{document}